\documentclass[a4paper]{jpconf}
\usepackage{graphicx}

\usepackage{epsfig}
\usepackage{amssymb,amsmath,bm}

\newcommand\f[2]{\frac{#1}{#2}}

\def\be{\begin{equation}}
\def\ee{\end{equation}}
\def\bea{\begin{eqnarray}}
\def\eea{\end{eqnarray}}

\def\as{{\alpha_s}}

\def\R{\mathrm{Re}}
\def\la{\langle}
\def\ra{\rangle}

\def\mR{\mu_{\mathrm{R}}}
\def\mF{\mu_{\mathrm{F}}}

\def\ep{\epsilon}
\def\cm{\mathcal{M}}

\begin{document}

\title{General formulation of the sector-improved residue subtraction}

\author{David Heymes}

\address{Institute for Theoretical Particle Physics and Cosmology,
  RWTH Aachen University, Sommerfeldstr. 16, 52074 Aachen, Germany}

\ead{dheymes@physik.rwth-aachen.de}

\begin{abstract}
The main theoretical tool to provide precise predictions for scattering
 cross sections of strongly interacting particles is perturbative QCD. 
Starting at 
next-to-leading order (NLO) the calculation suffers 
from unphysical IR-divergences that cancel in the final result. At NLO
 there exist general subtraction algorithms to treat these divergences 
during a calculation. Since the LHC demands for more precise theoretical 
predictions,  general subtraction methods at next-to-next-to-leading order
(NNLO) are needed.\\   
This proceeding outlines the four-dimensional formulation of the sector improved 
residue subtraction. The subtraction scheme STRIPPER and in particular
its extension to arbitrary multiplicities is explained. Therefore, 
it furnishes a general framework for the calculation of NNLO cross
sections in perturbative QCD.
\end{abstract}


\section{Introduction}
We are interested in predicting the hadronic cross section, which is
known to factorize into parton distribution functions and the
partonic cross section
\be
\sigma_{h_1h_2}(P_1, P_2) = \sum_{ab} \iint_0^1 \mathrm{d}x_1
\mathrm{d}x_2 \, f_{a/h_1}(x_1, \mF) \,
f_{b/h_2}(x_2, \mF) \, \hat{\sigma}_{ab}(x_1P_1, x_2P_2;
\, \as(\mR), \, \mR, \, \mF) \; .
\ee
The summation runs over initial state partons $\{a, b\}$,
i.e. massless quarks and gluons. The parton distribution function
$f_{a/h_1}(x_1,\mF)$ can be understood as the probability density for
finding parton $a$ inside hadron $h_1$ carrying the momentum
$p_1=x_1 P_1$. Parton distribution functions are non-perturbative
objects and have to be determined experimentally.\\  
In contrast, the partonic cross section $\hat{\sigma}_{ab}$ can be
calculated using perturbative QCD. Including terms
up to next-to-next-to-leading order, its expansion in the
strong coupling $\as$
reads
\be
\hat{\sigma}_{ab} = \hat{\sigma}^{(0)}_{ab} + \hat{\sigma}^{(1)}_{ab}
+ \hat{\sigma}^{(2)}_{ab} \; .
\ee
The leading order contribution is known as the Born approximation and
reads 
\be
\hat{\sigma}^{(0)}_{ab} = \hat{\sigma}^{\mathrm{B}}_{ab} =
\f{1}{2\hat{s}} \f{1}{N_{ab}} \int \mathrm{d} \bm{\Phi}_n \, \la
\cm_n^{(0)} | \cm_n^{(0)} \ra \, \mathrm{F}_n
\; ,
\ee
where $n$ is the number of final state particles and
$\mathrm{d}\bm{\Phi}_n$ the phase space measure. The measurement
functions $F_n$ defines the infrared safe observable and prevents $n$
massless partons from becoming soft or collinear. The $i$-loop matrix
element is denoted by $| \cm_n^{(i)} \ra$. For details of the
notation we refer to \cite{Czakon:2014oma}.\\
Beyond leading order, we decompose the cross section
according to the number of particles in the final state. At
next-to-leading order (NLO) we have
\be
\label{eq:NLO}
\hat{\sigma}^{(1)}_{ab} = \hat{\sigma}^{\mathrm{R}}_{ab} + 
\hat{\sigma}^{\mathrm{V}}_{ab} + \hat{\sigma}^{\mathrm{C}}_{ab} \; ,
\ee
with
\be
\hat{\sigma}^{\mathrm{R}}_{ab} =
\f{1}{2\hat{s}} \f{1}{N_{ab}} \int \mathrm{d} \bm{\Phi}_{n+1} \, \la
\cm_{n+1}^{(0)} | \cm_{n+1}^{(0)} \ra \, \mathrm{F}_{n+1} \; , \quad
\hat{\sigma}^{\mathrm{V}}_{ab} = \f{1}{2\hat{s}} \f{1}{N_{ab}} \int
\mathrm{d} \bm{\Phi}_n \, 2 \R \, \la \cm_n^{(0)} | \cm_n^{(1)} \ra \,
\mathrm{F}_n \; .
\ee
Starting at this order, separate contributions suffer from soft and collinear
(infrared) divergences. They appear as poles in the regulator $\ep$ after
setting the space-time dimension to $d=4-2\ep$. We distinguish between  
explicit virtual poles, that emerge in the one-loop matrix element $|
\cm_n^{(1)} \ra$ of the virtual contribution
$\hat{\sigma}^{\mathrm{V}}_{ab}$, and real poles, that appear after
integrating the phase space of the additional parton of the real
contribution $\hat{\sigma}^{\mathrm{R}}_{ab}$. All poles cancel in the
sum \eqref{eq:NLO}.\\
At next-to-next-to leading order (NNLO) we get
\be
\label{eq:NNLO}
\hat{\sigma}^{(2)}_{ab} = \hat{\sigma}^{\mathrm{RR}}_{ab} +
\hat{\sigma}^{\mathrm{RV}}_{ab} + \hat{\sigma}^{\mathrm{VV}}_{ab} +
\hat{\sigma}^{\mathrm{C1}}_{ab} + \hat{\sigma}^{\mathrm{C2}}_{ab} \; ,
\ee
where
\be
\label{eq:NNLOEXP}
\begin{gathered}
\hat{\sigma}^{\mathrm{RR}}_{ab} =
\f{1}{2\hat{s}} \f{1}{N_{ab}} \int \mathrm{d} \bm{\Phi}_{n+2}
\, \la \cm_{n+2}^{(0)} | \cm_{n+2}^{(0)} \ra \, \mathrm{F}_{n+2} \; ,
\\  \hat{\sigma}^{\mathrm{RV}}_{ab} =
\f{1}{2\hat{s}} \f{1}{N_{ab}} \int \mathrm{d} \bm{\Phi}_{n+1} \, 2 \R
\, \la \cm_{n+1}^{(0)} | \cm_{n+1}^{(1)} \ra \, \mathrm{F}_{n+1} \; ,
\\[0.2cm] \hat{\sigma}^{\mathrm{VV}}_{ab} =
\f{1}{2\hat{s}} \f{1}{N_{ab}} \int \mathrm{d} \bm{\Phi}_n \, \Big( 2
\R \, \la \cm_n^{(0)} | \cm_n^{(2)} \ra + \la \cm_n^{(1)} |
\cm_n^{(1)} \ra \Big) \, \mathrm{F}_n \; .
\end{gathered}
\ee
The double-real contribution $\hat{\sigma}^{\mathrm{RR}}_{ab}$
contains two additional massless partons in the final state that can
become unresolved and lead to poles in $\ep$ after the phase space
integration is performed. The real-virtual contribution
$\hat{\sigma}^{\mathrm{RV}}_{ab}$ consists of the one-loop amplitude 
integrated over the $n+1$ particle phase space. In addition to
virtual poles of the one-loop matrix element, it develops real poles by
integrating the phase space of one unresolved particle. The
double-virtual contribution $\hat{\sigma}^{\mathrm{VV}}_{ab}$ contains
only explicit virtual poles in the two-loop amplitude and the squared one-loop
amplitude. The sum of all contributions in \eqref{eq:NNLO} is finite.
\footnote{For simplicity collinear counterterms
  $\hat{\sigma}^{\mathrm{C}}_{ab}$,
  $\hat{\sigma}^{\mathrm{C1}}_{ab}$ and
  $\hat{\sigma}^{\mathrm{C2}}_{ab}$ are not mentioned in this
  discussion. For details see \cite{Czakon:2014oma}.}
\\
In general it is not possible to perform phase space integrations
analytically. Furthermore, to be able to compare a predicition with
experimental data the phase space integration should be 
implemented in a flexible Monte-Carlo software to adapt the observable
to the experimental setup easily.\\
Subtraction methods at NLO have been established to handle infrared
singularities before numerical integrations are
performed. Catani-Seymour subtraction \cite{Catani:1996vz} and FKS
subtraction \cite{Frixione:1995ms} are commonly used schemes.\\
At NNLO, subtraction schemes become more involved. Antenna
subtraction \cite{GehrmannDeRidder:2005cm} and $q_T$ - subtraction
\cite{Catani:2007vq} are the most advanced proposals and 
have already been applied to $e^+e^-\to 3\,\mathrm{jets}$
 \cite{GehrmannDeRidder:2007jk}, $pp\rightarrow
 2\,\text{gluonic jets}$ \cite{Currie:2013dwa,Currie:2014upa},
 Higgs production \cite{Catani:2007vq} and vector boson pair production
 \cite{Cascioli:2014yka,Gehrmann:2014fva} and other non-trivial
 examples \cite{Abelof:2014fza,Chen:2014gva}.\\
In these schemes poles cancel analytically. We illustrate this by
taking the real and virtual contribution of a NLO cross section in
\eqref{eq:NLO}. The real-radiation cross section is made integrable in
four dimensions by a suitable subtraction term that mimics the
behaviour of the squared matrix element in the singular limits. Adding
the subtracted term back and integrating it analytically over the unresolved
one-particle phase space provides poles that cancel the poles of the
virtual contribution. Finally, both phase space integrals are
numerically integrable in four dimensions.\\
At NNLO the procedure is similar: Subtraction terms for real
unresolved particles are introduced to render the phase space
integrable. Analytically integrated subtraction terms cancel the
explicit poles of virtual contributions.\\
Here we present STRIPPER (SecToR ImProved PhasE space for Real
radiation), a NNLO subtraction scheme that is completely numerical and
avoids cumbersome analytic integrations. The scheme was introduced in
\cite{Czakon:2010td} and generalized to arbitrary final states in
\cite{Czakon:2014oma}. It has been first applied to top-quark pair
production \cite{Czakon:2011ve}, and subsequently  to other processes
of low multiplicity:
Higgs $+$ jet \cite{Boughezal:2013uia},
charmless bottom quark decay \cite{Brucherseifer:2013cu},
top quark decay \cite{Brucherseifer:2013iv}, single top quark
production \cite{Brucherseifer:2014ama},
muon decay \cite{Caola:2014daa} and Z decay \cite{Boughezal:2011jf}.\\
The scheme was initially formulated using conventional
dimensional regularization (CDR), where momenta and spin
degrees of freedom of resolved and unresolved particles are treated in
$d=4-2\ep$ dimensions. Unresolved particles are either virtual particles of
loop contributions or real-radiated particles that can become soft or
collinear. In contrast to analytic subtraction schemes, momenta of
resolved particles are explicitly parameterized in $d>4$ dimensions.
The explicit dimension increases as the multiplicity of the final
state rises, e.g. for top-quark pair production already five
dimensions have been parameterized explicitly. It turns out that
STRIPPER in CDR is not applicable for high multiplicities. In
addition, tree-level Matrix elements, that appear in subtraction
terms, have to be provided to several powers in $\ep$. 
Available software only provides them up to $\ep^0$.\\
Thus, it has been necessary to reformulate the scheme in 't
Hooft-Veltman regularization (HV), where momenta and spin degrees of
freedom of resolved particles are four-dimensional.\\
In this proceeding, we explain the general idea of STRIPPER in order
to obtain a Laurent series in $\ep$ for
$\hat{\sigma}^{\mathrm{RR}}_{ab}$, where each coefficient can be
calculated numerically. Afterwards, we shortly point out how 
it can be reformulated in HV to provide a self-contained
subtraction scheme for NNLO calculations. The detailed description is
to be found in \cite{Czakon:2014oma}.      


\section{STRIPPER}
The subtraction scheme STRIPPER is an algorithmic method to extract real
singularities of different contributions in \eqref{eq:NNLO}. Each part
will be given as a Laurent series in $\ep$, where each coefficient has
been calculated numerically. The final result, after summing the
different parts, is finite.\\
We outline the method for $\hat{\sigma}^{\mathrm{RR}}_{ab}$ as given in
\eqref{eq:NNLOEXP}. Since it contains two additional, potentially
unresolved, partons, the phase space integral has the most complicated
infrared structure.\\
First, we split the phase space into double-collinear and
triple-collinear sectors. In a triple-collinear sector singularities
are generated as three specific partons become collinear to each other
and/or two of them soft. In a double-collinear sector singularities
emerge as two specific pairs of partons become collinear and/or
two of them soft.\\
In a next step, we parameterize the collinear particles in each
sector separately using energies and angles. We illustrate the
parametrization on the basis of a triple-collinear sector,  
where the three particles are in the final state: The reference
momentum indicating the triple-collinear direction is
denoted by $r^{\mu}$. The momenta of the unresolved
partons are named $u_1^{\mu}$ and $u_2^{\mu}$. Each momentum is
parameterized by its energy and a $(d-1)$-dimensional unit vector in
spherical coordinates
\be
r^\mu \equiv r^0 \begin{pmatrix} 1 \\ \bm{\hat{q}_1} \end{pmatrix}
\; , \quad 
u_1^\mu \equiv u_1^0 \begin{pmatrix} 1 \\ \bm{\hat{u}_1} \end{pmatrix}
\; , \quad
u_2^\mu \equiv u_2^0 \begin{pmatrix} 1 \\ \bm{\hat{u}_2} \end{pmatrix}
\;.
\ee
The unresolved particles' energies are rescaled by their maximal value
$E$, $u_i^0={\hat \xi}_iE$, for $i=1,2$. The soft limit is approached as
${\hat \xi_1}\to 0$ and/or ${\hat \xi}_2\to 0$.
Using rotations in $(d-1)$-dimensions, the
scalar products between the three given momenta take the following
form
\be
\bm{\hat{u}_1} \cdot \bm{\hat{r}} = \cos\theta_1 = 1-2{\hat \eta}_1 \; , \quad
\bm{\hat{u}_2} \cdot \bm{\hat{r}} = \cos\theta_2 = 1-2{\hat \eta}_2 \; , \quad
\bm{\hat{u}_1} \cdot \bm{\hat{u}_2} = \cos\theta_1
\cos\theta_2 + \cos\phi_2 \sin\theta_1\sin\theta_2
\; .
\ee
\begin{figure}[h]
 \includegraphics[scale=.75]{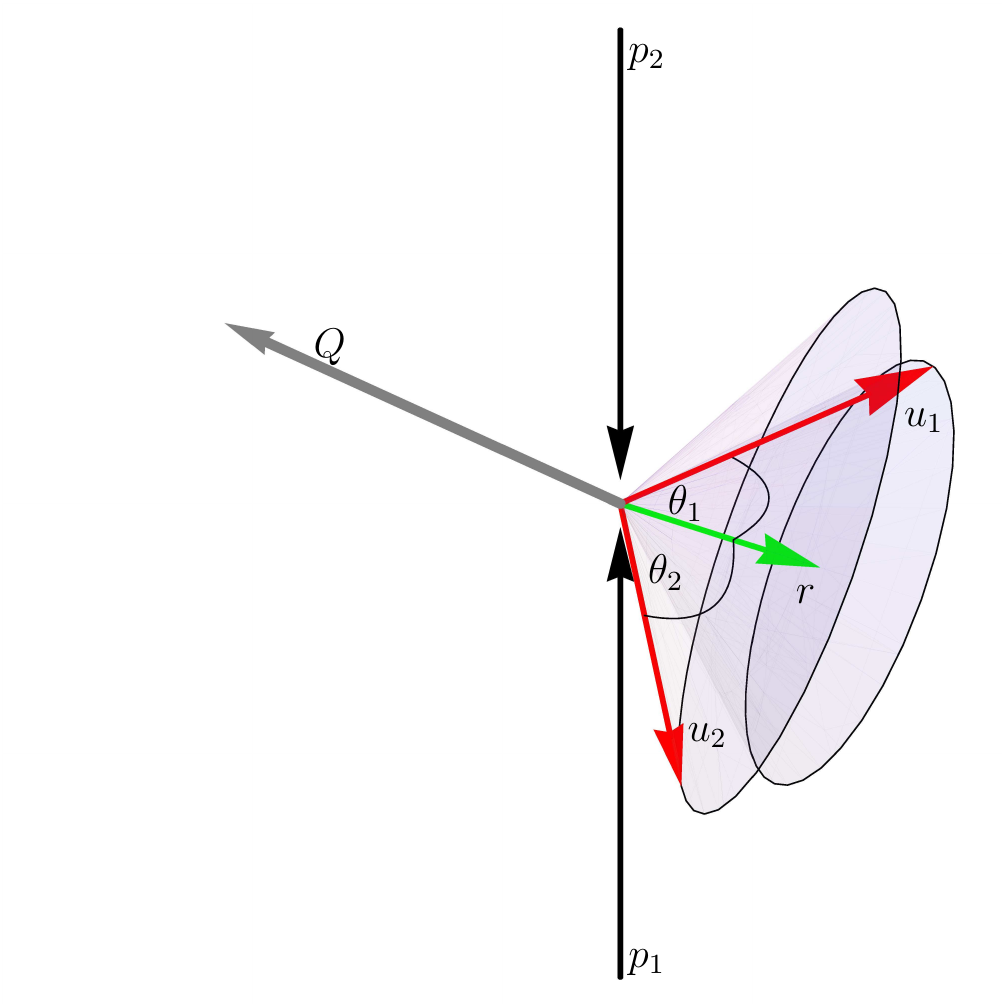}\hspace{2pc}%
\begin{minipage}[b]{14pc}
\caption{\label{fig:Parametrization} Parametrization of a
  triple-collinear sector, where the three partons are in the final
  state. The reference momentum is $r$, the unresolved momenta are
  denoted by $u_1$ and $u_2$. The sum of  momenta of remaining
  resolved particles is $Q$.}
\end{minipage}
\end{figure}
This parametrization is shown in figure \ref{fig:Parametrization}.
The limit of ${\hat \eta}_i$ at zero indicates the collinear limit
of one of the unresolved partons and the reference
parton. $\bm{\hat{u}_1}$ is collinear to $\bm{\hat{u}_2}$ when 
${\hat \eta}_1 = {\hat \eta}_2$ and $\phi_2=0$. 
A non-linear transformation of $\phi_2$ to $\zeta$,
\be
\phi_2 \rightarrow \phi_2({\hat \eta}_1,{\hat \eta}_2,\zeta) \; , 
\ee
then ensures that $\phi_2$ always vanishes as ${\hat\eta}_1={\hat \eta}_2$.
Accordingly, all possible collinear limits are indicated only by two
variables: ${\hat \eta}_1$ and ${\hat \eta}_2$. Furthermore,
if we find a formulation in HV, meaning that all other momenta and
the reference momentum are four-dimensional, at most six dimensions
are needed to parameterize all possible scalar products
consistently.\\
At this point, four physical variables $\{{\hat \eta}_1,{\hat \eta}_2,{\hat
  \xi}_1,{\hat \xi}_2\}$ parameterize all
possible soft and collinear limits in a given sector. Additional
sector decompositions \cite{Binoth:2000ps} in those variables
factorize all possible overlapping singularities that appear at
NNLO. In practice this amounts to split each sector again. For example
double-soft overlapping singularities are disentangled using  
\be
\label{eq:Decomp}
1=\theta\left(\hat{\xi}_1-\hat{\xi}_2\right)+
\theta\left(\hat{\xi}_2-\hat{\xi}_1\right).
\ee
The decomposition of the phase space due to soft overlapping
singularities is sufficient to factorize all possible limits in a
double-collinear sector. In a triple-collinear sector the phase space
is split into five additional sectors to factorize collinear and
soft-collinear overlapping singularities. This splitting is
accompanied by a transition from physical variables
$\{{\hat\eta}_1,{\hat \eta}_2,{\hat  \xi}_1,{\hat \xi}_2\}$ to
corresponding sector variables $\{ \eta_1, \eta_2,  \xi_1,\xi_2\}$. \\
Finally, the full double-real radiation cross section can be written as a
sum over different decomposed triple- and double-collinear sectors 
\be
{\hat
  \sigma}^{\mathrm{RR}}_{ab}=\sum_{S}{\hat
  \sigma}^{\mathrm{RR},S}_{ab}\; .
\ee
Each contribution has the following form
\be
{\hat
  \sigma}^{\mathrm{RR},S}_{ab}=\int\limits_0^1
\mathrm{d}\xi_1\mathrm{d}\xi_2\mathrm{d}\eta_1\mathrm{d}\eta_2 \frac{F_S\left(\xi_1,
  \xi_2, \eta_1,
  \eta_2\right)}{\xi_1^{1-b_1\epsilon}\xi_2^{1-b_2\epsilon}\eta_1^{1-b_3\epsilon}\eta_2^{1-b_4\epsilon}}.
\ee
The function $F_S\left(\xi_1,  \xi_2, \eta_1,
\eta_2\right)$ is finite in the limit of vanishing
arguments. All appearing singularities are factorized in
the sector variables. Poles are extracted by an iterative usage of
the plus distribution in each variable
\be
\label{eq:Plus}
\int_0^1\mathrm{d}x\,\frac{f\left(x\right)}{x^{1-b\epsilon}}=\frac{f\left(0\right)}{b\epsilon}+\int_0^1\mathrm{d}x\,\frac{f\left(x\right)-f\left(0\right)}{x^{1-b\epsilon}} \;.
\ee
We obtain a Laurent series in $\ep$ and all coefficients are
calculated numerically.
The described procedure is process independent, since it is possible
to use the known universal infrared-limits of QCD amplitudes for the
subtraction terms in \eqref{eq:Plus}. 
%

\section{Formulation in the 't Hooft-Veltman regularization scheme}
The above procedure is carried out in CDR, i.e. resolved and
unresolved momenta are $d$-dimensional. To get a transition to the 't
Hooft-Veltman regularization scheme, we have to identify unresolved
particles in different contributions of \eqref{eq:NNLO}: Either one particle is
unresolved, what we call the single-unresolved contribution, or two
particles are unresolved, what we call the double-unresolved
contribution. By adding process independent correction terms to the
single-unresolved contribution and subtracting them from the
double-unresolved contribution we make sure that the two contributions
are finite separately. Now we are able to identify unresolved and
resolved particles unambiguously. Since each contribution is finite we
set the momenta of resolved particles to $d=4$ dimensions, the mistake
is of order $\ep$. This makes sure that explicit parameterized
dimensions are limited to six for arbitrary multiplicities.
Setting also their spin degrees of freedom to
the physical dimensions allows us to use matrix elements at order
$\ep^0$. \\
To avoid contractions between $d$-dimensional vectors of unresolved
momenta and four-dimensional matrix elements, we replace
spin-correlated splitting functions in collinear subtraction terms by
their azimuthal averaged counterparts.\\
This procedure leads to a fully general subtraction algorithm for NNLO calculations.   
%


\section{Summary and conclusion}

In this proceeding, we presented the general formulation of
STRIPPER, a subtraction scheme to calculate cross sections at NNLO in
perturbative QCD for arbitrary multiplicities. The full subtraction
scheme, completely described in \cite{Czakon:2014oma}, can readily
implemented in a computer code.  


\section*{Acknowledgments}

\noindent

This research was supported by the German Research Foundation (DFG)
via the Sonderforschungsbereich/Transregio SFB/TR-9 ``Computational
Particle Physics''. The author thanks the organizers of ACAT 2014
for remission of the conference fee. 


\end{document}